# SYNTA: A novel approach for deep learning-based image analysis in muscle histopathology using photo-realistic synthetic data

**Leonid Mill[1, 2, *], Oliver Aust[3], Jochen A. Ackermann[3], Philipp Burger[3], Monica Pascual[3], Katrin Palumbo-Zerr[3], Gerhard Krönke[3], Stefan Uderhardt[3], Georg Schett[3], Christoph S. Clemen[4,5], Rolf Schröder[6], Christian Holtzhausen[6], Samir Jabari[6], Andreas Maier[2] and Anika Grüneboom[7, †]**

[1] MIRA Vision Microscopy GmbH, 73054 Eislingen / Fils, Germany

[2] Pattern Recognition Lab, Friedrich-Alexander University Erlangen-Nürnberg (FAU), 91058 Erlangen, Germany

[3] Department of Internal Medicine 3 - Rheumatology and Immunology, Universitätsklinikum Erlangen, Friedrich-Alexander University Erlangen-Nürnberg (FAU), 91054 Erlangen, Germany

[4] Institute of Aerospace Medicine, German Aerospace Center (DLR), Cologne, Germany

[5] Institute of Vegetative Physiology, Medical Faculty, University of Cologne, Cologne, Germany

[6] Department of Neuropathology, Universitätsklinikum Erlangen, Friedrich-Alexander University Erlangen-Nürnberg (FAU), 91054 Erlangen, Germany

[7] Leibniz-Institut für Analytische Wissenschaften - ISAS - e.V., 44139 Dortmund, Germany

[*] Lmill@mira.vision, [†] anika.grueneboom@isas.de

## Abstract

Artificial intelligence (AI), machine learning, and deep learning (DL) methods are becoming increasingly important in the field of biomedical image analysis. However, to exploit the full potential of such methods, a representative number of experimentally acquired images containing a significant number of manually annotated objects is needed as training data. Here we introduce SYNTA (synthetic data) as a novel approach for the generation of synthetic, photo-realistic, and highly complex biomedical images as training data for DL systems. We show the versatility of our approach in the context of muscle fiber and connective tissue analysis in histological sections. We demonstrate that it is possible to perform robust and expert-level segmentation tasks on previously unseen real-world data, without the need for manual annotations using synthetic training data alone. Being a fully parametric technique, our approach poses an interpretable and controllable alternative to Generative Adversarial Networks (GANs) and has the potential to significantly accelerate quantitative image analysis in a variety of biomedical applications in microscopy and beyond.

## Introduction

Technical developments in the field of diagnostic and high throughput imaging are accompanied by an increasing generation of large-scale image data sets [1]–[3]. Especially for applications in biomedical imaging and digital pathology, which often rely on quantitative data analysis [4]–[6] an accurate and manual evaluation of such big datasets is almost impossible within an acceptable time frame. While fully automated, deep learning (DL)-based methods showed to be a reliable solution for this problem [7]–[9], such approaches follow the supervised learning paradigm [7], [8] thus requiring labor-intensive manual annotation of large datasets. Additionally, as such networks are typically trained towards specialized datasets, they often suffer from poor generalization on similar but different data [10]. This so-called *domain shift* is generally caused by changes in e.g., the tissue and staining type, sample preparation protocol or the data acquisition hardware [10], [11].

While in recent years several approaches [5], [9] have been introduced to successfully tackle the generalization problem, these methods still rely on significant amounts of human-labeled images [8], [13]–[15]. On the other hand, synthetic training data do not require manual annotation, they can be used to increase the diversity and reduce biases in datasets [16], [17], and are already being used for various computer vision tasks [17]–[19] including image segmentation in microscopy [16], [20]–[22]. Yet, in the context of complex biomedical images in digital pathology, to date only unsupervised image synthesis methods based on Generative Adversarial Networks [23], [24] (GANs) are reported to produce realistic results [25]–[31]. However, being purely data driven methods, GANs are typically considered as black boxes that cannot be interpreted nor allow to fully control its output. Additionally, it is not possible to automatically extract ground truth (GT) labels from GAN-generated images, which however, are essential to train Convolutional Neural Networks (CNNs) on image segmentation tasks.

In this work, we introduce SYNTA (synthetic data) as a novel approach for the photo-realistic generation of highly complex synthetic biomedical images as training data for a state-of-the art segmentation network in the context of digital pathology. As a proof of concept, we performed a comprehensive quantitative assessment of images from skeletal muscle tissue sections derived from desmin knock-out (DKO) mice [32]–[34] a well-established animal model for human autosomal-recessive desminopathies with lack of desmin [35]. For a diagnostic evaluation of skeletal muscle specimens, number, size and form of muscle fibers give important basic information allowing classification of the sample in either healthy or diseased [36]. Manually segmenting muscle fibers or measuring the fiber diameter is a time-consuming process, usually limited to small sample sizes (~400 fibers) [37], [38] and is prone to human variance. We show that our approach, which does not rely on GANs but is solely based on interpretable parametric texturing and computer graphics, can be used to train a robust fiber segmentation network to predict expert-level results on previously unseen but real-world data. Furthermore, we demonstrate the model's superior generalization capabilities compared to a network that was trained on real data. Finally, we analyze over 208,000 muscle fibers from a real-world dataset based on our approach, with the fully automated analysis taking approximately 5 minutes per entire tissue section. Given the large number of quantitatively analyzed muscle fibers, we show that even a single fiber feature (diameter) is sufficient to distinguish between control and pathology induced tissue samples.

## Results

### Biomedical image segmentation using photo-realistic synthetic data.

Following the supervised learning paradigm, to train a state-of-the art CNN (U-Net [6], [39]) segmentation network, a considerable number of images need to be manually annotated (Fig. 1a). However, as sample preparation protocols and data acquisition hardware may vary during experiments and among institutions, the differences in the resulting datasets often lead to the *domain-shift* [10] causing poor segmentation performances of U-$Net_{real}$ on the new dataset B (Fig. 1b). To reliably address this problem, in general a refinement of the model on the new dataset is needed by adding a significant number of representative and manually annotated new samples to the training data.

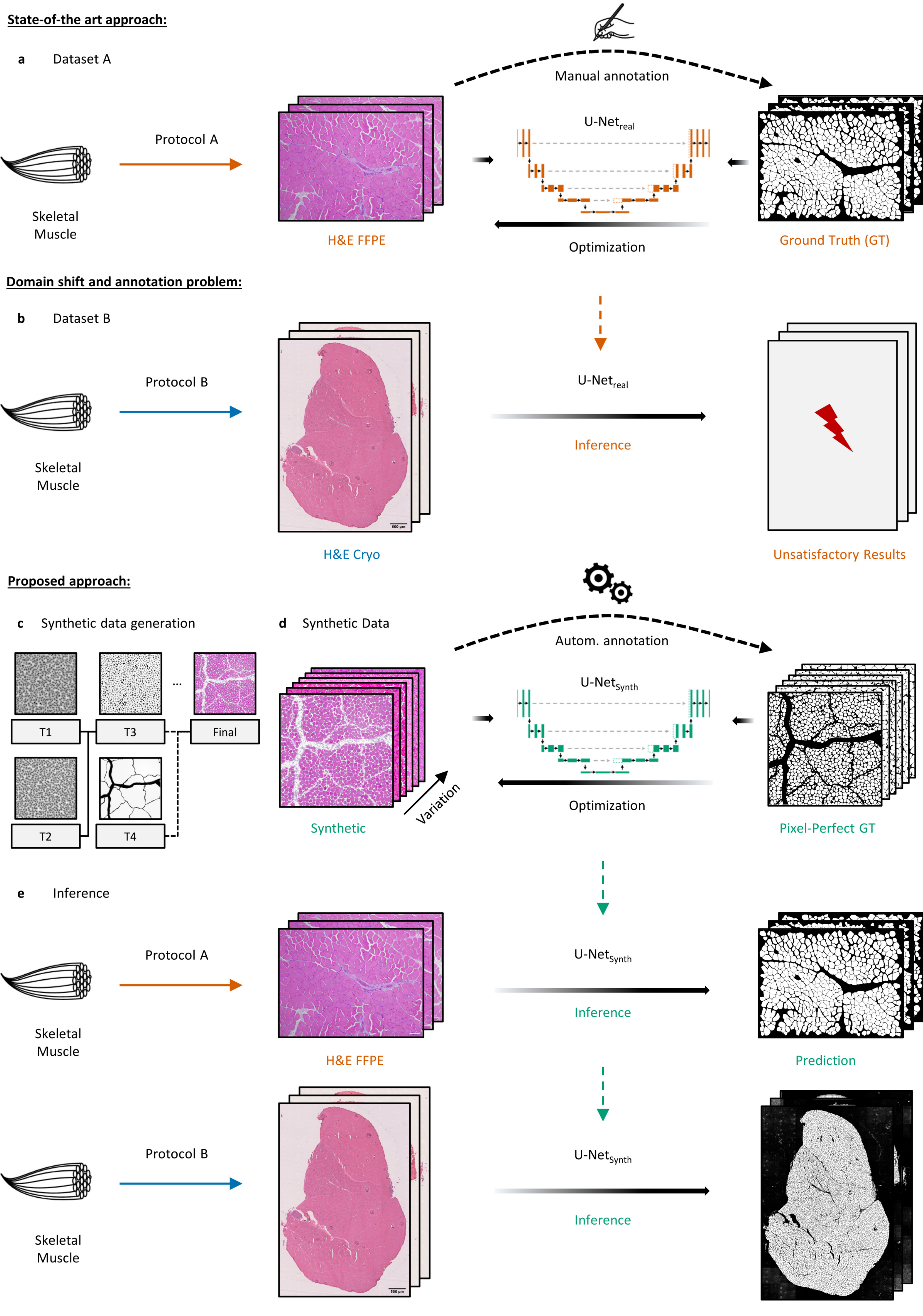

**Fig. 1 | Illustration of the state-of-the art approach for the segmentation of biomedical image data, the problems that are associated with this approach and the proposed methodology (SYNTA) using photo-realistic synthetic training data. a,** The classical approach for DL based biomedical image segmentation requires the manual annotation of a (specialized) dataset to train a model ($\text{U-Net}_{real}$) on the fiber segmentation task. **b,** As sample preparation protocols or the acquisition hardware can change during experiments or among institutions, the difference in distributions between the two datasets A and B can lead to the *domain shift* resulting in poor generalization of $\text{U-Net}_{real}$ across the datasets. **c,** Illustration of the SYNTA data generation pipeline. Pre-defined parametric textures (T1-T4) were used to create a simulation pipeline that imitates the inherent features (fiber shapes, tissue artifacts, staining variance, etc.) of H&E-stained normal murine skeletal muscle brightfield images. **d,** Based on the simulation, a photo-realistic and highly diverse synthetic dataset containing over 74,000 muscle fibers with its pixel-perfect GT labels was generated and used to train a segmentation model ($\text{U-Net}_{synth}$) on the synthetic data only. **e,** Although solely trained on synthetic data, $\text{U-Net}_{synth}$ was not only able to predict expert-level segmentation on previously unseen real-world data, but it also generalized well on both datasets A (H&E-stained sections of formalin-fixed paraffin-embedded skeletal muscle tissue (H&E FFPE)) and B (H&E-stained sections of cryo-preserved skeletal muscle tissue (H&E Cryo)).

In this work we propose an alternative solution to this problem with SYNTA. We used real-world images as visual reference and hand-crafted a pipeline of parametric textures (Fig. 1c, T1-T4) in a 3D computer graphics software (the open-source software Blender [40]) to generate the inherent features (fiber shapes, tissue artifacts, staining variance, etc.) present in H&E-stained skeletal muscle brightfield images (see Supp. Fig. 1a). Once the SYNTA simulation pipeline was implemented, this fully interpretable simulation approach allowed us to automatically render a highly diverse photo-realistic synthetic dataset containing over 74,000 muscle fibers and its respective pixel-perfect labels. We then trained a U-Net ($\text{U-Net}_{synth}$) on the synthetic data only (Fig. 1d). After the training procedure, $\text{U-Net}_{synth}$ was not only able to perform well on previously unseen real-world data, but the diversity within the synthetic dataset also resulted in $\text{U-Net}_{synth}$ to generalize well across the datasets A and B, resulting in expert-level fiber segmentation quality for both real datasets (Fig. 1e). A refinement of $\text{U-Net}_{synth}$ on real data was not necessary.

**Photo-realistic synthetic biomedical images.**

We based our simulation on the features of real skeletal muscle reference images to achieve the highest possible degree of photo-realism for the synthetic images (see Methods section). In this context, Fig. 2a shows a visual comparison of a real skeletal muscle image (left) of dataset A (H&E FFPE, H&E-stained sections of formalin-fixed paraffin-embedded normal murine skeletal muscle tissue) and a synthetic image generated using the proposed SYNTA concept (right). At this point we want to emphasize, that our simulation technique does not require any kind of real image data as input nor is it based on GANs.

Both, the real (left) and synthetic image (right) contain the typical features of microscopic brightfield images of H&E-stained cross-sections derived from skeletal muscle. These include e.g., muscle fiber shapes, peripheral nuclei, the endomysium, and perimysium components of the connective tissue as well as tissue artifacts. Additionally, due to the parametric nature of our simulation, we were also able to introduce a considerable spectrum of realistic variations within the synthetic images in terms of the staining representation, fiber properties (densities, sizes and shapes), the position and size of the connective tissue as well as the density and position of peripheral and central nuclei (Fig. 2b, Supp. Fig. 1).

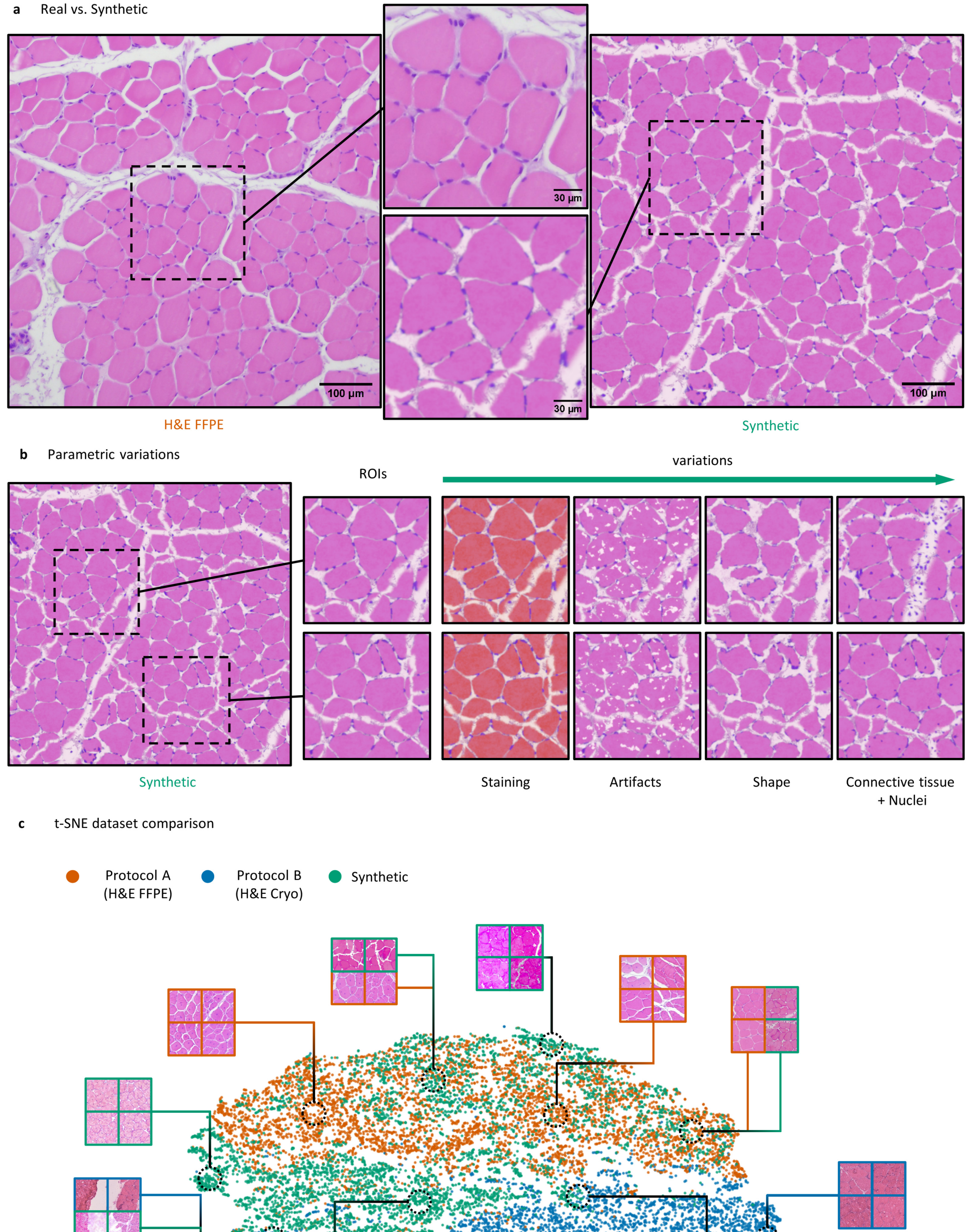

**Fig. 2 | Direct comparison of synthetic and real data. a,** Exemplary real image of a H&E FFPE (dataset A) stained normal murine muscle tissue acquired using a brightfield microscope (left) in comparison to a photo-realistic synthetic image (right) generated using our simulation approach. **b**, Example parametric variations within a single synthetic image. The pipeline allows to control every feature of the synthetic image to create realistic variations. This includes a diversity of the staining, sizes, shapes, positions and densities of muscle fibers, nuclei, and connective tissue components. **c,** t-SNE plot comparing the three datasets. Each data point represents a randomly extracted image patch of the size 256 x 256 pixels. In total, 6,000 image patches were extracted for the t-SNE visualization for each dataset, respectively.

To assess the similarity between the two real-world (H&E FFPE and H&E Cryo) and the synthetic datasets, the structure of the datasets was visualized in Fig. 2c by using a t-distributed stochastic neighbor embedding [41] (t-SNE). For the visualization, 6,000 image patches of the size 256 x 256 pixels were extracted from each dataset, respectively. The distribution of the H&E FFPE (orange) and H&E Cryo (blue) data points show well separated clusters with only little overlap. In contrast, the synthetic data points (green) are scattered across the embedding space while showing intersections with both real-world datasets. This suggests an overall high degree of similarity between synthetic and real data points. However, the relatively strong overlap of H&E FFPE and synthetic samples may indicate a higher degree of similarity between these two datasets in comparison to the H&E Cryo data.

### Segmentation performance on the H&E FFPE data.

We next wanted to assess whether the synthetic dataset was photo-realistic enough such that a state-of-the art deep segmentation network was able to operate on real data while being trained on synthetic data only. To test this hypothesis, we trained a U-Net (U-$Net_{synth}$) on $n=120$ synthetic images containing over 74,000 muscle fibers and compared its segmentation performance to a U-Net (U-$Net_{real}$), which was trained on real H&E FFPE images ($n=13$) with over 6,200 expert-annotated fibers. Due to the comparably limited number of real H&E FFPE images, U-$Net_{real}$ was trained in a leave-one-out cross-validation setup to assess its general segmentation performance. As additional benchmark, we compared the U-Net predictions to the performance of the generalist cytoplasm model of Cellpose [8], which was pre-trained on a variety of cell and non-cell images, including images of muscle fibers and containing over 70,000 segmented objects [8]. As post processing routine for the U-Net predictions, we additionally implemented a fully automated Fiji [42] macro toolset. A detailed description of the post processing steps, and other features of the toolset can be found in Supp. Fig. 2-4.

The qualitative and quantitative segmentation results for the H&E FFPE datasets are displayed in Fig. 3. For the qualitative results, two exemplary test patches are shown in Fig. 3a. While the manual GT annotations are visualized as black contour lines, the fiber contours of the model predictions are indicated by dashed yellow lines.

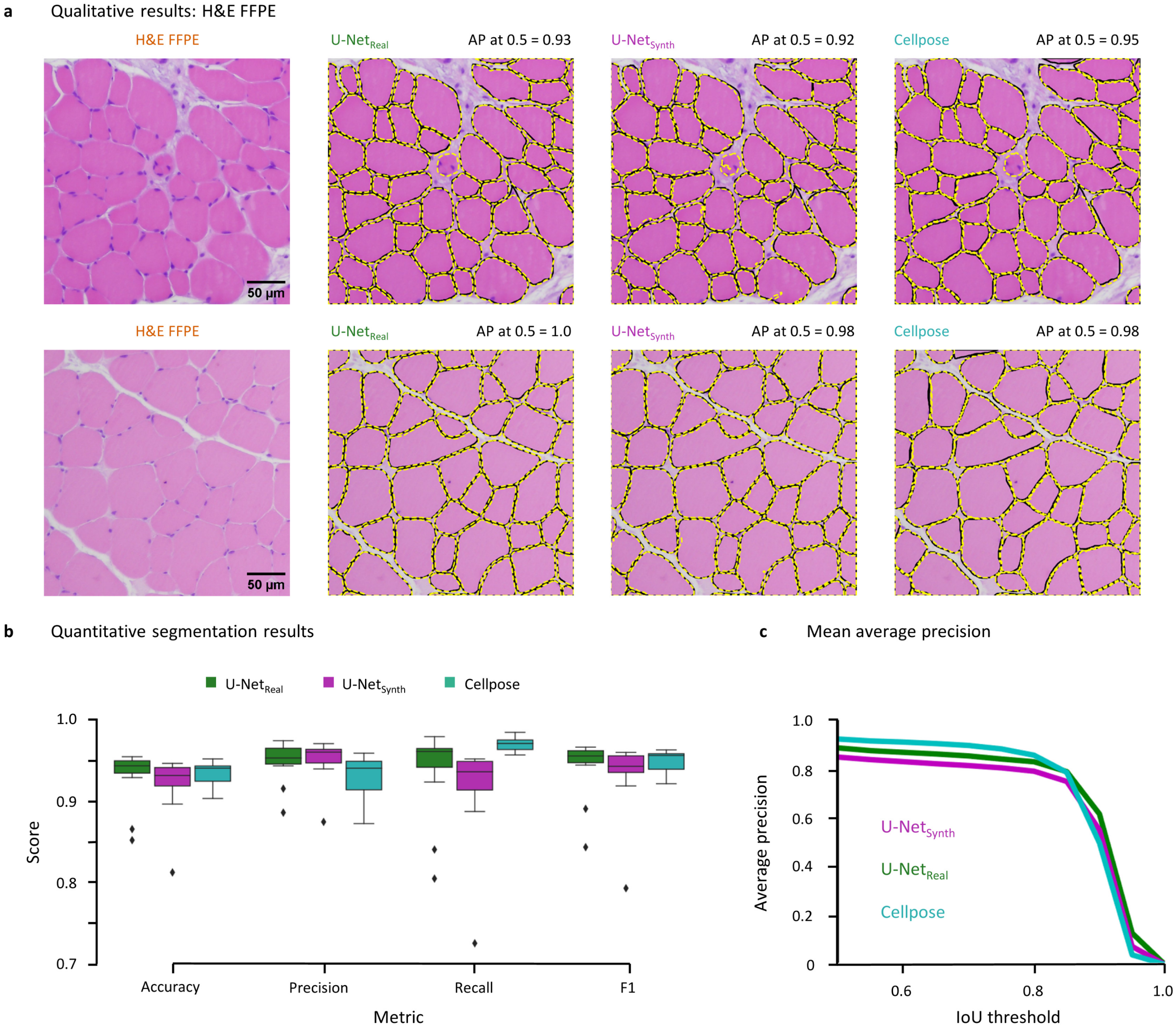

**Fig. 3 | Segmentation results on the H&E FFPE data (dataset A). a,** Qualitative segmentation results of U-Net$_{real}$, U-Net$_{synth}$ and Cellpose on two example test images. The black fiber contours indicate the manual GT annotations, the dashed yellow lines denote the model predictions. The average precision (AP) at an IoU of 0.5 is given for each model above the respective images. **b,** Comparison of the quantitative segmentation results for the H&E FFPE dataset based on the accuracy, precision, recall and F1 metrics. **c**, Instance segmentation performance based on the mean AP for a IoU threshold in the range of [0.5, 1] in steps of 0.05.

For both test images (top and bottom row), all model predictions match the GT contour lines almost perfectly. These results are also reflected by the average precision (AP) at an Intersection over Union (IoU) threshold of 0.5, which is denoted above each prediction. The quantitative results for the H&E FFPE dataset are visualized in Fig. 3b, c. Across all evaluation metrics U-Net$_{synth}$ (accuracy = 0.92, precision = 0.95, recall = 0.91, F1 = 0.93) showed a similar performance compared to U-Net$_{real}$ (accuracy = 0.93, precision = 0.95, recall = 0.94, F1 = 0.94) and Cellpose (accuracy = 0.93, precision = 0.93, recall = 0.97, F1 = 0.95). Also, the generalist model of Cellpose (AP$_{0.5}$ = 0.93) marginally outperformed U-Net$_{real}$ (AP$_{0.5}$ = 0.89) and U-Net$_{synth}$ (AP$_{0.5}$ = 0.85) according to the AP at the commonly used IoU threshold of 0.5. The performance of U-Net$_{synth}$ was in line with the results of U-Net$_{real}$ and Cellpose, although U-Net$_{synth}$ was trained on synthetic data only.

**Segmentation performance on the H&E Cryo data derived from wild-type and desmin knock-out mice.**

Based on the conclusive results of U-Net$_{synth}$ on H&E FFPE data, we tested for the assumption that the proposed approach of using synthetic training data alone would lead to a more robust DL model. Hence, we further compared the generalization capabilities of both U-Nets (U-Net$_{real}$ and U-Net$_{synth}$) on the H&E Cryo dataset. However, as no large-scale manual annotations were available for this dataset, we conducted an expert survey to grade the predictions of both models. In a first step, two neuropathology experts (P1 and P2) predefined 156 analytically relevant regions of interest (ROIs) in $n_{WSI}=52$ whole slide images (WSI). Afterwards, for each predefined ROI, the two experts evaluated three different segmentation masks: the quality of a fiber segmentation mask "*Segmentation*" containing diagnostically relevant and non-relevant fibers, a shape filtered segmentation mask "*Shape Filter*" that should contain only diagnostically relevant fibers by removing fibers according to its specific shape features (e.g. based on their diameter) and the quality of the connective tissue segmentation that is based on the non-shape-filtered fiber segmentation (Fig. 4a). For the survey, P1 and P2 graded each segmentation mask according to an ordinal scale (i.e., "very bad", "bad", "intermediate" (int.), "good" and "very good"). In total, 156 ROIs were evaluated by P1 and P2, for each model (U-Net$_{real}$ and U-Net$_{synth}$) and each category "Segmentation", "Shape Filter" and "Connective Tissue", respectively.

Fig. 4b provides a qualitative comparison of the segmentation results of U-Net$_{real}$ and U-Net$_{synth}$ for an expert-predefined ROI in images of H&E-stained sections of cryo-preserved soleus/gastrocnemius skeletal muscle tissue derived from homozygous desmin knock-out mice (DKO Hom, myopathic murine muscle) and wild-type siblings (WT healthy murine muscle), respectively. The top row visualizes the post processed fiber segmentation masks, the second row displays the shape-filtered fiber masks, while the bottom row shows the connective tissue segmentations. Both models, U-Net$_{real}$ and U-Net$_{synth}$, were able to predict meaningful segmentations. However, the qualitative results show that U-Net$_{synth}$ is not only equal to but indeed outperforms U-Net$_{real}$ on the fiber segmentation task. It generalized well on the H&E Cryo images providing precise and very accurate fiber segmentations. In contrast, U-Net$_{real}$ often failed to find distinct fiber contours resulting in over-segmented fibers, which were subsequently removed by the shape filter leading to considerably more missing fibers in this mask. Additionally, U-Net$_{real}$ failed to handle freezing artifacts well while U-Net$_{synth}$ on the other hand demonstrated to be more robust against such artifacts, leading to more accurate fiber segmentations (Fig. 4b, ROI WT). Consequently, the connective tissue segmentation of U-Net$_{synth}$ also shows more accurate results compared to U-Net$_{real}$.

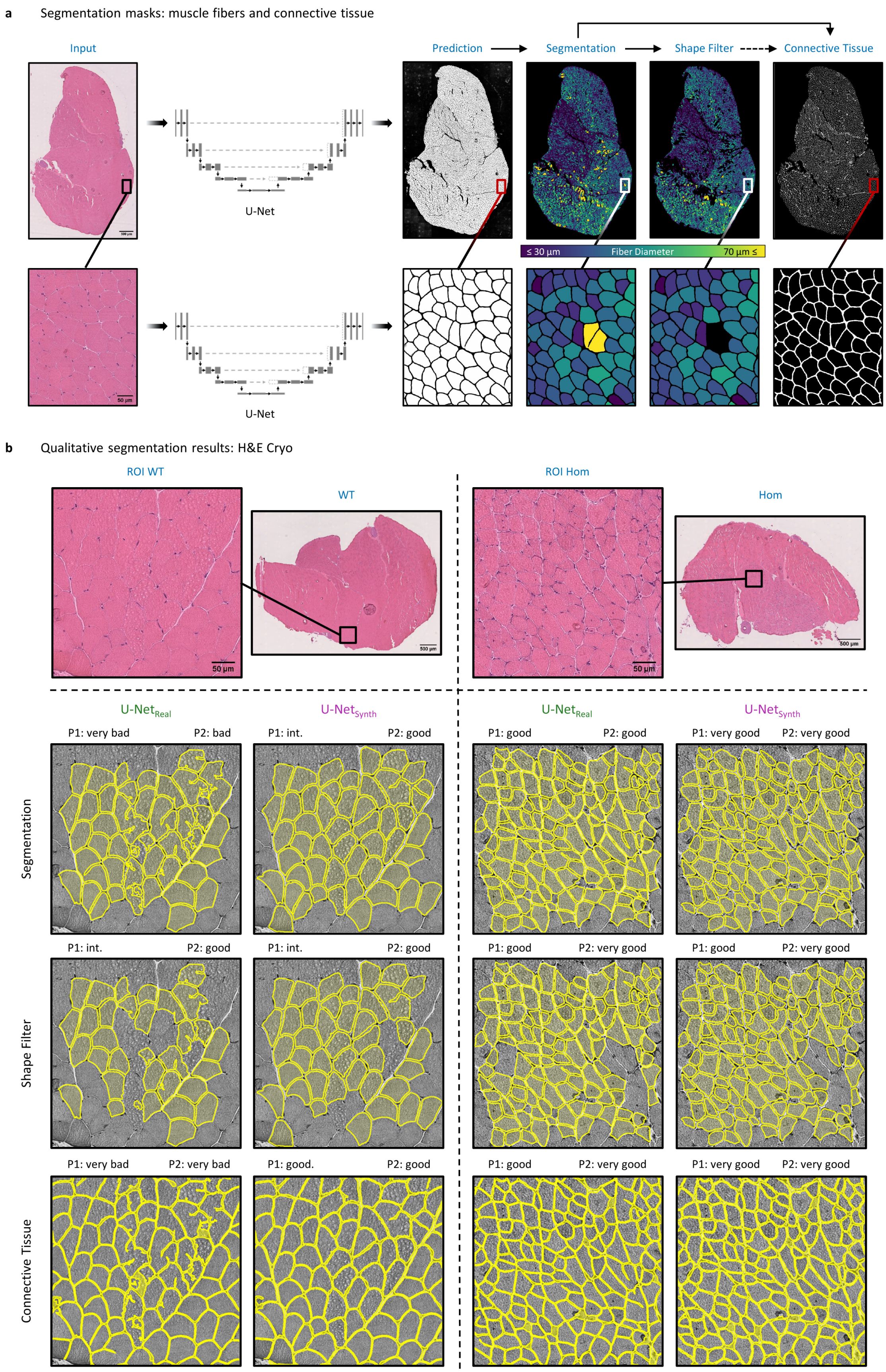

**a** Segmentation masks: muscle fibers and connective tissue

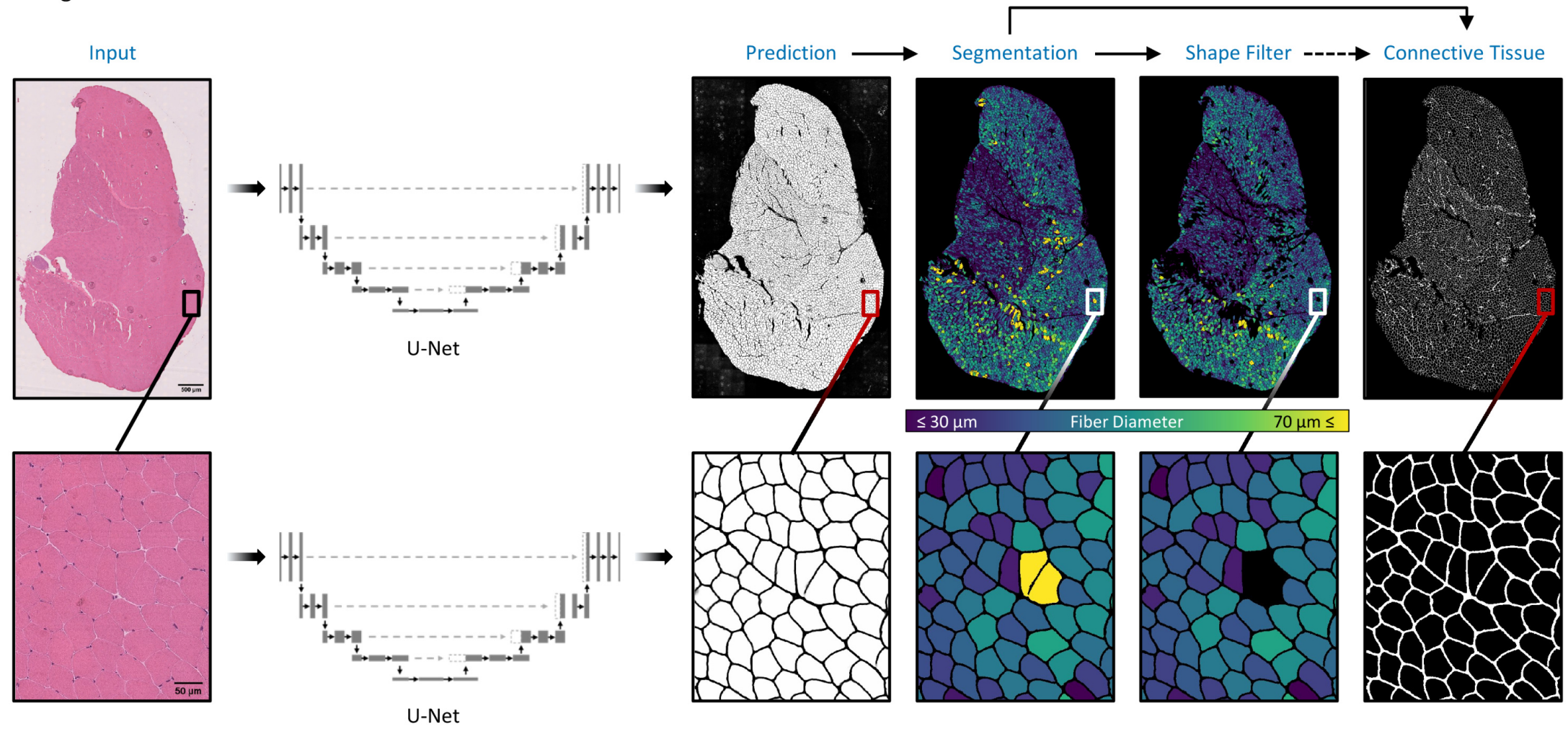

**b** Qualitative segmentation results: H&E Cryo

ROI WT

WT

ROI Hom

Hom

50 µm

500 µm

50 µm

500 µm

U-Net$_{Real}$ U-Net$_{Synth}$ U-Net$_{Real}$ U-Net$_{Synth}$

Segmentation

P1: very bad P2: bad | P1: int. P2: good | P1: good P2: good | P1: very good P2: very good

Shape Filter

P1: int. P2: good | P1: int. P2: good | P1: good P2: very good | P1: good P2: very good

Connective Tissue

P1: very bad P2: very bad | P1: good. P2: good | P1: good P2: very good | P1: very good P2: very good

**Fig. 4 | Segmentation results on the H&E Cryo data. a,** Model prediction based segmentation masks ("Segmentation", "Shape Filter" and "Connective Tissue") for an exemplary selected H&E Cryo whole slide image (WSI) of normal soleus/gastrocnemius skeletal muscle tissue derived from a wild-type mouse (WT). **b,** To assess the model performance of U-Net$_{real}$ and U-Net$_{synth}$ on the unlabeled H&E Cryo dataset, the three segmentation masks were rated by two experts (P1 and P2) according to an ordinal scale. The top row exemplary shows two H&E Cryo expert pre-defined ROIs with the respective qualitative segmentation results of U-Net$_{real}$ and U-Net$_{synth}$. The ROIs refer to normal and myopathic soleus/gastrocnemius skeletal muscle tissue derived from wild-type siblings (WT) and homozygous desmin knock-out mice (DKO Hom), respectively. The expert ratings for the ROIs are denoted above the predictions for each category ("Segmentation", "Shape Filter" and "Connective Tissue"), respectively.

The quantitative results for the survey are provided in Fig. 5, which shows a diverging stacked bar chart [43] based on the P1 and P2 expert ratings. According to the survey results, U-Net$_{synth}$ predictions were consistently rated more favorably in all three categories. The results also show that predictions from U-Net$_{synth}$ were in general considered as "Intermediate", "Good" or "Very good" while in comparison the predictions from U-Net$_{real}$ were rated "Bad" or "Very bad" more frequently. Overall, the U-Net$_{synth}$ received significantly better ratings than the U-Net$_{real}$ in all categories examined.

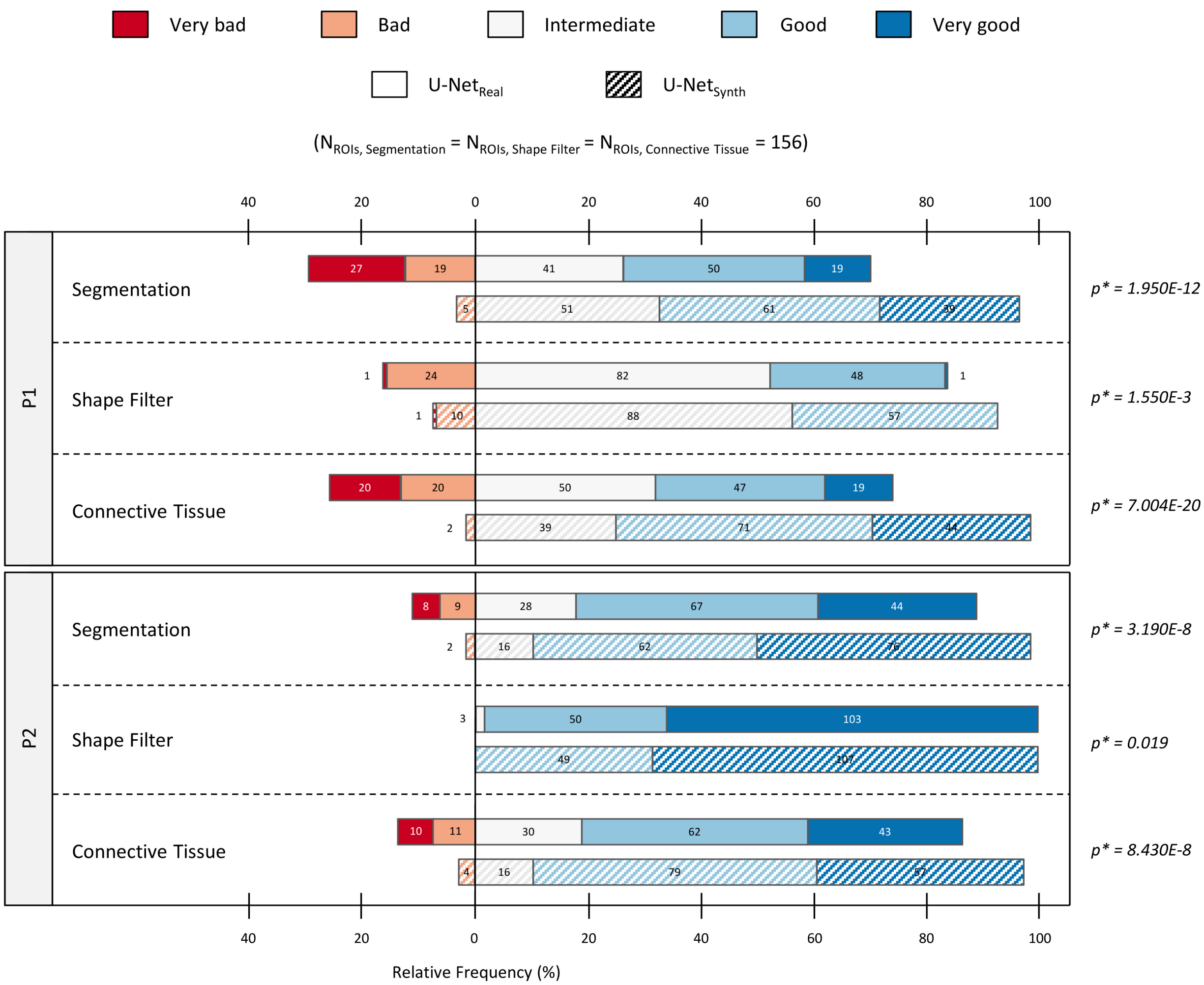

**Fig. 5 | Expert survey results and the exclusion of regions with severe artifacts for the subsequent H&E Cryo dataset analysis.** Diverging stacked bar chart showing the quantitative results of an expert survey evaluating the segmentation performances of U-Net$_{real}$ and U-Net$_{synth}$ on the unlabeled H&E Cryo dataset. Based on an ordinal scale each expert (P1 and P2) ranked the models' predictions for 156 expert pre-defined ROIs for the categories "Segmentation", "Shape Filter" and "Connective Tissue", respectively. The absolute number and relative frequency of ROIs graded as "very bad" or "bad" are shown to the left of the zero line. The absolute number and relative frequency of predictions, which were ranked as "intermediate",

"good" or "very good" are visualized to the right of the zero line. Furthermore, for each category ("Segmentation", "Shape Filter" and "Connective Tissue") the p-values of a paired Student's t-test are provided.

This means that according to the P1 and P2 evaluations, U-Net$_{synth}$ significantly outperformed U-Net$_{real}$ in all three categories, producing in general more accurate fiber and connective tissue segmentation masks. Based on these encouraging results, we used U-Net$_{synth}$ to analyze the complete H&E Cryo dataset.

**In-depth H&E Cryo dataset analysis.**

To ensure an accurate analysis of the entire H&E Cryo dataset, we annotated specific regions within those sections that contained freezing artifacts (Type I) and artificial loosening (Type II) and subsequently removed such regions from the "Shape Filter" and the "Connective Tissue" segmentation masks (Supp. Fig. 5). From overall nine tissue samples (murine soleus/gastrocnemius skeletal muscle packages, 5 WT (healthy, wild-type siblings) and 4 DKO Hom (abnormal/mypathic, homozygous desmin knock-out) we obtained 52 WSI or sections (24 WT and 28 Hom sections) and analyzed a total of *208,270* muscle fibers (Fig. 6a). The mean computation time for the automated analysis per section was ~5 minutes (~2 minutes for the U-Net$_{synth}$ predictions and 3 minutes for post processing and feature extraction). Desmin knock-out (DKO) mice [32]–[34] are a well-established animal model for human skeletal muscle myopathies and cardiomyopathies [44]. Skeletal muscle tissue derived from these mice shows a myopathic pattern comprising an increased variation of fiber diameters and content of connective tissue that is most prominent in the soleus muscle [45]. To validate the performance of our approach in a biological context, we investigated whether it was possible to differentiate between WT and DKO Hom samples solely based on the features extracted from the U-Net$_{synth}$ "Shape filter" and "Connective Tissue" segmentation masks. Note that the "Shape Filter" mask was used as the basis for the dataset analysis as the shape filter was optimized towards containing diagnostically relevant muscle fibers only (see Supp. Fig. 3 and Supp. Fig. 4). Additionally, to investigate whether the phenotype was present differently in distinct muscle regions we manually annotated the musculus soleus (M. soleus) region for a subsequent differentiation of the analysis results between the whole section as well as the M. soleus and M. gastrocnemius parts.

The fiber diameters were analyzed considering either the whole section, the M. soleus only or the M. gastrocnemius only. In all three cases, WT sections could clearly be linearly separated from the DKO Hom sections (Fig 6b). The fiber diameter distributions showed, as expected, a distinct shift towards smaller muscle fibers in DKO Hom thus leading to an increased diameter variance (see Supp. Fig. 6). In addition to the fiber diameter analysis, we further investigated the disease phenotype by addressing the local thickness of the connective tissue in either the M. soleus or M. gastrocnemius parts of the whole section.

**a** Extracted fiber features

| Sample type | # Samples | # Sections | Fiber features ($\mu \pm \sigma$) | | | Total fiber count |
|---|---|---|---|---|---|---|
| | | | Area [µm²] | Diameter [µm] | Fiber count | |
| WT | 4 | 24 | 1,567.5 ± 783.5 | 36.6 ± 11.0 | 5552.5 ± 459.6 | 133,259 |
| DKO Hom | 5 | 28 | 1,450.0 ± 1095.9 | 33.6 ± 13.7 | 2679.0 ± 670.2 | 75,011 |

**b** Fiber diameter distribution

**c** Connective tissue local thickness distribution

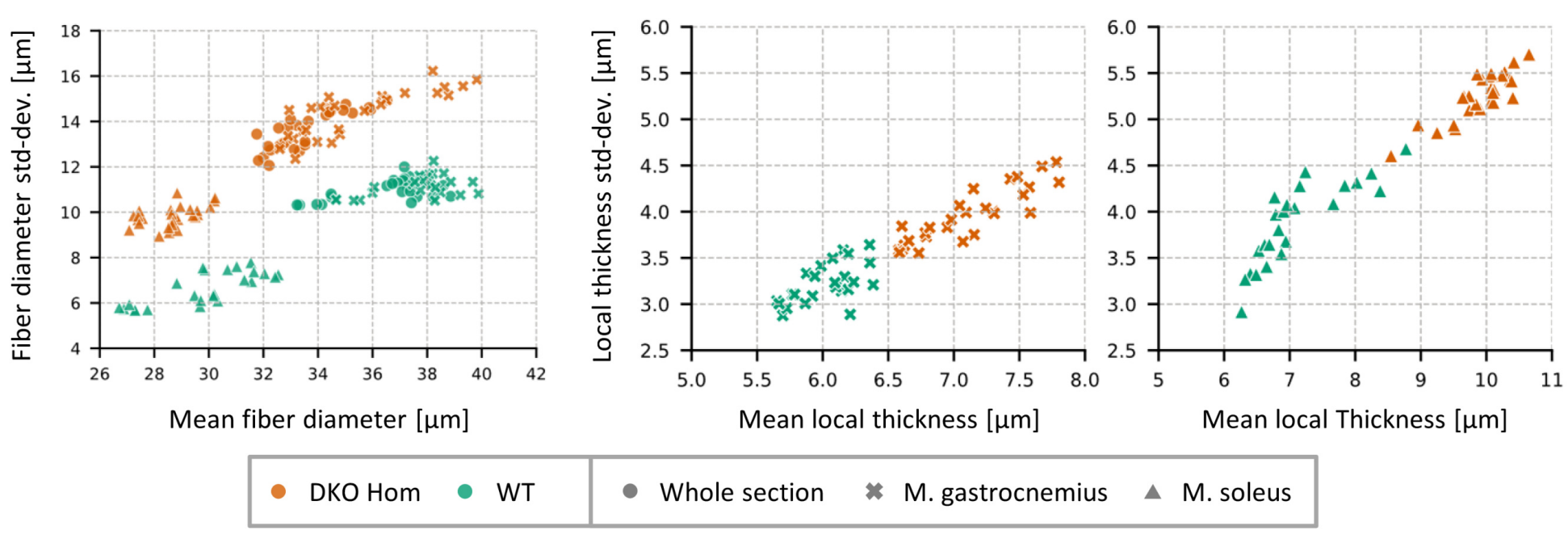

**d** Abnormal vs. normal: muscle fibers and connective tissue

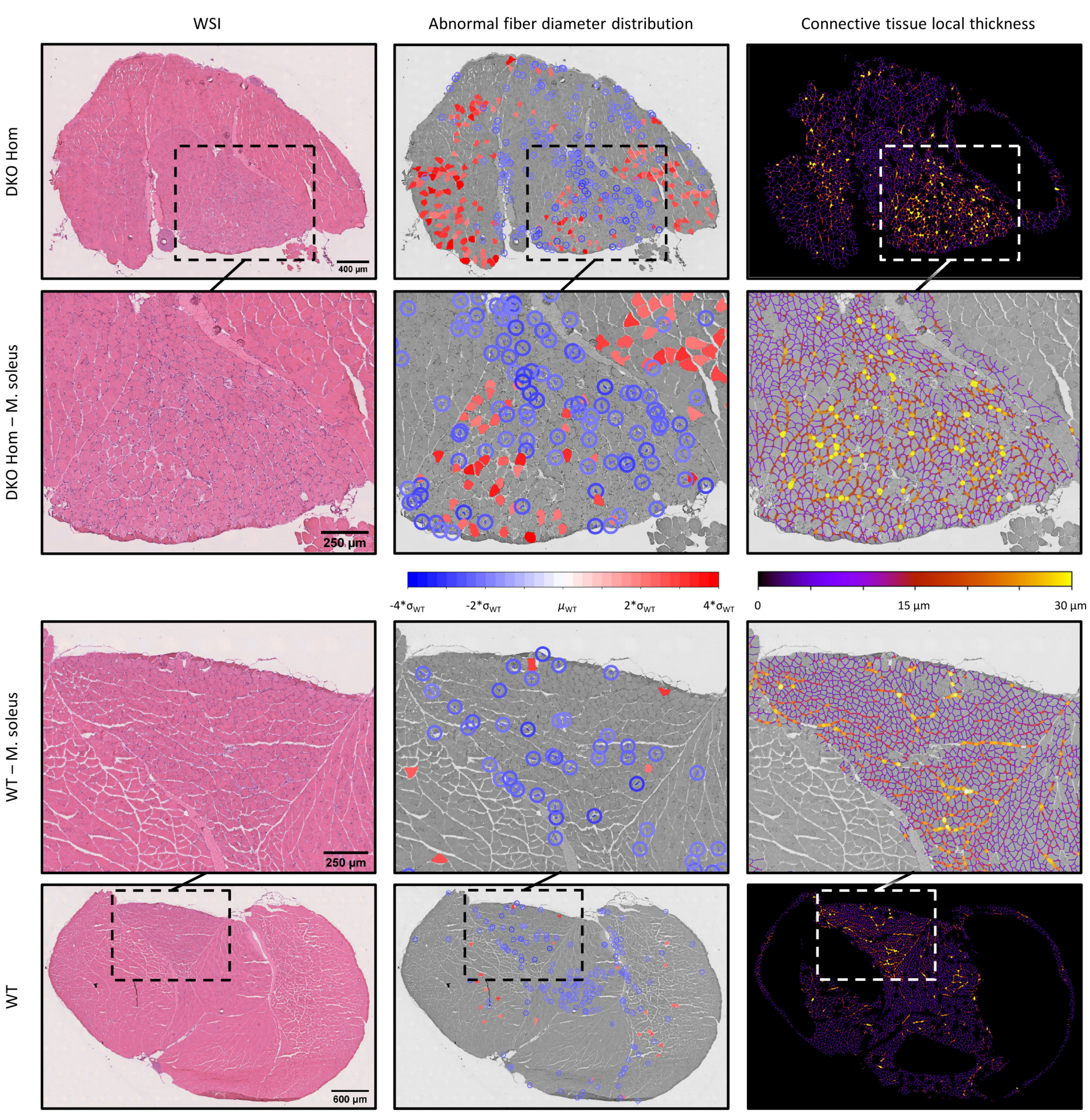

**Fig. 6 | Analysis of the H&E Cryo dataset based on the U-Net$_{synth}$ predictions. a,** Extracted fiber features from the H&E Cryo dataset. **b,** Scatter plot of the fiber diameter distribution showing the mean fiber diameter (in µm) in relation to the fiber diameter standard deviation (std-dev.) (in µm). Each scatter point represents either a WT (healthy, wild-type sibling, in green) or a DKO Hom (abnormal/myopathic, homozygous desmin knock-out, in orange) section. Fiber diameters were analyzed considering either the whole section (circles), the M. soleus only (triangles) or the M. gastrocnemius only (crosses). **c,** Scatter plots of the connective tissue local thickness distribution for each tissue section considering solely the M. soleus (right) or the M. gastrocnemius (left) parts. **d,** Abnormal/myopathic vs. normal muscle fiber distribution and connective tissue based on the abnormality analysis.

Fig. 6c shows the resulting scatter plots that display the mean local thickness (in µm) of the connective tissue for the M. soleus and M. gastrocnemius regions of each section, respectively. The results demonstrate that it is also possible to distinguish between WT or DKO Hom samples considering solely the local thickness of the connective tissue.

Based on the significant number of analyzed WT muscle fibers, we next addressed the possibility to depict the general disease phenotype by assessing only local, specific regions in single sections by identifying abnormal fibers. In the context of the abnormality analysis, we defined a muscle fiber as abnormal if the difference of its diameter to the WT mean fiber diameter $\mu_{WT}$ was larger than $|2\sigma|$. Note that due to the differences of M. soleus and M. gastrocnemius in terms of mean fiber diameters (M. soleus fibers are in general smaller than M. gastrocnemius fibers, also see Supp. Fig. 6), M. soleus fibers were compared against $\mu_{\text{WT, M. soleus}}$, while M. gastrocnemius fibers were compared against $\mu_{\text{WT, M. gastrocnemius}}$.

Fig. 6d displays all identified abnormal fibers in exemplary chosen DKO Hom and WT sections. For visualization purposes, abnormal fibers that were significantly smaller (blue) than $\mu_{WT}$ were encircled for a better visual identification. The analysis revealed that fibers of a small size in DKO Hom sections are mostly located in the M. soleus region with only a small number of fibers with increased diameter. In the M. gastrocnemius compartment close to the M. soleus part also many small fibers could be found. The above observed increase of the local thickness of the connective tissue was present in these areas with abnormal fiber size values. On the other hand, in the remaining part of the M. gastrocnemius both small and increased fibers were identified.

## Discussion

Here we introduced SYNTA as a novel concept for the generation of highly complex and photo-realistic synthetic histopathology images as training data for a state-of-the art segmentation network (U-Net). We validated our approach on the fiber segmentation in brightfield images of H&E-stained skeletal muscle sections for two different datasets. We showed that the quality, the complexity, and the high degree of photo-realism of the synthetic dataset was sufficient to train a U-Net for the precise segmentation of muscle fibers in real-world images. Furthermore, a comparison of a U-Net trained on real data (U-Net$_{real}$) with a U-Net trained on synthetic data only (U-Net$_{synth}$) revealed that the performance of both models was not only similar, but U-Net$_{synth}$ was able to generate expert-comparable fiber segmentation results on previously unseen real data. Additionally, our experiments demonstrated the superior generalization capability of U-Net$_{synth}$, which significantly outperformed U-Net$_{real}$ on a second (but also real) dataset. We finally analyzed over 208,000 skeletal muscle fibers based on the expert-level performance of U-Net$_{synth}$ and showed, that it is possible to distinguish between healthy control and diseased (desmin knock-out myopathy) tissue samples solely based on a single fiber feature or the thickness of the connective tissue.

While in this work we could show that photorealistic rendering can be used to achieve remarkable results in the context of the segmentation of highly complex biomedical images, it is important to emphasize that it requires certain expertise in computer graphics to create realistic and simulation-based image data.

However, taking in mind the diversity and complexity of real-world biomedical images, our concept poses a suitable alternative to the financial costs and the time effort needed to acquire biological samples and collect representative real-world training datasets that contain a significant number of manually annotated objects. Using synthetic content generation provides the flexibility that is needed to incorporate realistic repeating image features and patterns, which are often observable in microscopic images (cells, nuclei, shapes, etc.), in interpretable and controllable simulations. Being a fully parametric approach, the proposed concept enables the integration of domain knowledge and human expertise directly in the data simulation process and thus the training data itself. This not only leads to more robust and unbiased DL models, but it also allows to incorporate specific image features that a DL model implicitly learns during training. Hence, we are confident that this approach poses an interpretable and controllable alternative to state-of-the art image synthesis methods such as GANs and is a first step to significantly accelerate quantitative image analysis in a variety of biomedical applications in microscopy and beyond.

## Methods

### Ethics declaration:

Health monitoring of all mice was done as recommended by the Federation of European Laboratory Animal Science Associations (FELASA). Mice were handled in accordance with the German Animal Welfare Act (Tierschutzgesetz) as well as the German Regulation for the protection of animals used for experimental or other scientific purposes (Tierschutz-Versuchstierverordnung). Male wildtype C57BL/6J mice were housed in isolated cages under germ free conditions in a standard environment with free access to water and food. Sample preparation was carried out in compliance with all ethical regulations for organ removals at the University of Erlangen and carried out as approved by the local animal ethic committees of the Regierung von Mittelfranken (reference number 55.2.2-2532-2-1073). Female homozygous desmin knock-out (DKO) mice B6J.129S2/Sv-*Des*$^{tm1Cba}$/Cscl [32]–[34] and their wild-type siblings were housed in isolated ventilated cages (IVC) under specific and opportunistic pathogen-free (SOPF) conditions in a standard environment with free access to water and food. All investigations were approved by the governmental office for animal care (Landesamt für Natur, Umwelt und Verbraucherschutz North Rhine-Westphalia (LANUV NRW), Recklinghausen, Germany (reference numbers 84-02.04.2014.A262 and 84-02.05.40.14.057)).

### H&E FFPE (formalin-fixed paraffin-embedded) sample preparation and staining protocol

Muscles of male C57BL/6J wildtype mice were isolated at 15 weeks of age, fixed for overnight in phosphate-buffered 4% formaldehyde (Roti-Histofix, Carl Roth GmbH) and stored in 70% Ethanol for several days. Muscles were embedded in paraffin and 2 µm sections were cut and stained for H&E.

Paraffin slides were dewaxed by applying Xylol two times for 5 minutes each at room temperature (RT). Then slides were immersed in 100% isopropyl alcohol for 2 minutes twice followed by two times immersion in 96% isopropyl alcohol for 2 minutes. Deionized water was used to wash the slices. Mayer*'s* hemalum solution (Merck 1.092490500) was mixed 1:10 with deionized water and slides were immersed in the resulting solution for 10 minutes at RT. Slides were washed with deionized water, followed by washing with 1% HCl-ethanol. Then slides were exposed to running tap water for 10 min at RT. Slides were immersed in Eosin solution (Sigma) consisting of 0,5% Eosin and 0,01% acetic acid in deionized water and left in this solution at RT for 15 min. Deionized water was used to wash the slides again, followed by immersion in an increasing ethanol series with a final brief exposure to EBE (Merck). Finally, slides were mounted using a xylol free mounting medium.

### H&E FFPE dataset (dataset A)

The H&E FFPE dataset consisted of 12 expert-annotated images of the size 4140 x 3096 pixels, containing in total 6204 labeled muscle fibers. The image acquisition was performed using a Zeiss Axio Lab.A1 laboratory microscope and the cellSens Entry Software (OLYMPUS, version 1.3) at 10x optical zoom (Zeiss N-Achroplan 10x/0,25 Ph1 ∞/- objective) and a resolution of 0.34 µm per pixels (2.933 pixels per µm).

## H&E Cryo sample preparation and staining protocol

Gastrocnemius/soleus muscle groups were dissected from female DKO Hom and healthy WT controls at age 28 – 34 weeks, embedded in Tissue-Tek OCT compound (Sakura Finetek), immediately frozen in liquid nitrogen-cooled isopentane, and cryostat sections of 6 µm thickness were collected on microscope slides, air-dried for 30 min, and used for H&E stains.

Cryosections were incubated in filtrated hematoxylin solution (Gill II, Epredia 6765007) for 2 minutes. Afterwards, slides were rinsed in room temperature tap water, then incubated in tap water for 2 minutes, and immediately transferred into an eosin staining solution (Bio Optica, 05-100003/L) for 30 seconds. For dehydration, slides were briefly incubated in isopropanol (1x 70%, 2x 96%, 2x 100%), and then transferred into xylene before mounting with a coverslip.

## H&E Cryo dataset (dataset B)

The H&E Cryo dataset consisted of 27 WSI images of varying image sizes, acquired using a Hamamatsu NanoZoomer S60 (C13210) whole slide scanner at 40x magnification with a scanning resolution of 0.22 µm per pixels. From the 27 WSI, we manually extracted 52 WSI (tissue sections) at a resolution of 0.6615 µm per pixels (1.5117 pixels per µm) using QuPath [46] and Fiji [42].

## Simulation of skeletal muscle bright-field microscopy images

The simulation of photo-realistic skeletal muscle bright-field microscopy images is based on texturing techniques, which allow to create fully parametric simulations through combinations of textures. This enables the generation of large randomized digital content by deterministic and human pre-defined parameters. During the content generation process, the texture parameters are sampled from a (pre-defined) range of values following specific design rules, such that the computed output of the pipeline resembles realistic variations. In this work, we used this technique in combination with the open-source 3D software Blender [40] (version 2.93) to render realistic microscopic images of skeletal muscle tissue. To do so, we first visually inspected the features (fiber and nuclei shapes and coloring, tissue artifacts, connective tissue etc.) of real-world reference images to gain sense of the diversity within such images. Afterwards, we used the rendering software to hand-craft a comprehensive texturing pipeline, which was mainly based on the Worley noise algorithm [47] in order to imitate a wide range of the observed variations (see Fig. 2b, Supp. Fig. 1). After the implementation of the pipeline, we pre-defined a set of deterministic parameter ranges to ensure that the generated results resemble a diverse but realistic variety of synthetic skeletal muscle images.

## Synthetic dataset

Based on the parametric texturing pipeline, we automatically generated 120 synthetic images of the size 2048 x 2048 pixels using the Python application programming interface of Blender. During this data generation process, various random variations were introduced (see Fig. 2b) to alter the appearance of the muscle fibers, nuclei, connective tissue (perimysium and endomysium) and the background. These included variations for the staining, size, shape, and distribution of fibers and nuclei as well as diverse changes for the connective tissue and tissue artifacts. For each rendered image, a corresponding ground truth fiber segmentation mask was automatically extracted from the pipeline. In total, the synthetic dataset contained approximately 74,000 muscle fibers with their respective pixel-perfect GT annotations. Based on fiber size comparisons of the synthetic dataset with real data, the pixel resolution distribution within the synthetic dataset was 0.79 ± 0.22 µm per pixel.

**Dataset comparison**

Three datasets (H&E FFPE, H&E Cryo and the synthetic dataset) were compared via t-SNE [41] as illustrated in Fig. 2. The t-SNE visualization is based on image features extracted by a VGG16 [48], which was pre-trained on the ImageNet [49] dataset. As input for the VGG16, we used 6000 random image patches of the size 256 x 256 pixels from each dataset. To further reduce the dimensionality of each feature vector, we applied a principal component analysis [50] (PCA) while covering 90% of its variance. Afterwards, t-SNE with a perplexity of 50, 1000 iterations, a learning rate of 50 and a Euclidean distance metric was used to visualize all data points. The t-SNE algorithm was applied using the scikit-learn [51] (version 0.24.0) Python machine learning framework.

**U-Net specifications and training**

Both U-Nets [39] (U-Net$_{real}$ and U-Net$_{synth}$) share the same three-class architecture predicting background, fibers and boundary pixels as segmentation masks, while only the fiber segmentation masks were considered for the analysis. The input of the networks was a three-channel (RGB) image. As network architecture, we used the specifications proposed by the original U-Net publication [39]. Additionally, we used Group Normalization [52] as normalization techniques during training and applied bilinear upscaling instead of the originally proposed deconvolutions in the decoder part of the network.

The training of both U-Nets (U-Net$_{real}$ and U-Net$_{synth}$) was performed on three-channel (RGB) image patches for 120 epochs using stochastic gradient descent and the Adam [53] optimizer. As training parameters, we used cross-entropy loss, learning rate decay with an initial learning rate of 0.001 and a mini-batch size of one with 150 iterations per epoch. The image patch size was set to 400 x 400 pixels. For data augmentation, random 90° rotations, additive noise and Gaussian blur was applied in combination with random brightness, contrast, saturation, and hue changes of an image patch. Before feeding images to the network, the images were normalized as additional pre-processing step. To strengthen the focus of the networks on the fiber borders, a pixel-wise weight map was computed on-the-fly during the training process following the proposed approach from ref [39]. In this context, we found that setting the $w_0$ parameter to a very high value of ~400 with $\sigma = 5$ pixels achieved the best results. To compensate for the high weighting of the border pixels, we additionally included a weighting for the fiber bodies by a factor of ~50 and additionally weighted background pixels by a factor of 10. Both networks (U-Net$_{real}$ and U-Net$_{synth}$) were trained using the same training parameters. The networks were implemented and trained using the PyTorch [54] machine learning framework (version 1.10.0). The implementation was based on ref [55].

U-Net$_{real}$ was trained on the complete H&E FFPE dataset with 12 manually expert-annotated images of the size 4140 x 3096 pixels. In total, the dataset contained 6204 labeled muscle fibers. As the synthetic data contained approximately 3.3x times more pixel data and ~12x times more annotated fibers (synth: 74221, H&E FFPE: 6204), U-Net$_{real}$ was trained in a leave-one-out cross-validation setup to assess its general segmentation performance. Thereby, early stopping was used to prevent the models from overfitting on the training data. In this regard, 9% of the training set served as validation data.

The training of U-Net$_{synth}$ was performed on the entire synthetic dataset using the same training parameter as used for U-Net$_{real}$. For the selection of the best performing U-Net$_{synth}$ model, we followed the model selection approach as proposed by Mill et al. [20]: after each epoch the model performed a segmentation on a real (H&E FFPE) validation image patch of the size 1000 x 1000 pixels. Afterwards, the segmentation performance was assessed qualitatively for each prediction. Based on this evaluation, the model with the best performance was chosen for the segmentation of all real (H&E FFPE and H&E Cryo) data.

### Benchmarking

The segmentation performance of U-Net$_{real}$ and U-Net$_{synth}$ was compared to the generalist cytoplasm model of Cellpose [8]. Cellpose is a state-of-the art generalist segmentation method which was trained on a dataset of varied cell images containing over 70,000 manually segmented objects, including images of muscle fibers. For the benchmarking on the H&E FFPE dataset, we used the official implementation from ref [8] with an object diameter size of 120 pixels and the default flow threshold of 0.4.

### Quantitative Evaluation on H&E FFPE data

The model performances were quantitatively evaluated on the H&E FFPE dataset using 6,204 manually annotated ground truth fiber masks as baseline. For each prediction the accuracy, precision, recall, and F1 score [56] was computed and averaged over all images.  Additionally, to assess the segmentation quality on a per-object basis, we quantified the overlap between a predicted and its corresponding GT fiber mask using the mean average precision [57] (AP) as instance segmentation metric. In this context, we provide the AP for varying IoU thresholds in the interval of [0.5, 1.0] in steps of 0.05 (see Fig. 3c).

### Expert survey

For the expert survey on the H&E Cyro dataset, three ROI per section have been defined by a pathology expert (one ROI in the M. soleus and two ROI within the M. gastrocnemius), resulting in total 152 ROIs for the experiment. The size of the ROI was set to 400 x 400 µm. For each ROI, the two pathology-experts (P1 and P2) ranked the segmentation masks "*Segmentation*", "*Shape Filter*" and "*Connective Tissue*" (see Fig. 4) of U-Net$_{real}$ and U-Net$_{synth}$, respectively. In this context, to reduce and eliminate experimental bias, the survey was conducted as a blinded experiment: for every section, the segmentation results of Net$_{real}$ and U-Net$_{synth}$ were randomly masked as segmentation from model 'A' or 'B'. To account for the pixel resolution differences among the training datasets of U-Net$_{real}$ (H&E FFPE) und U-Net$_{synth}$ (synthetic dataset), predictions of U-Net$_{real}$ on the H&E Cryo dataset were obtained at a pixel resolution of 0.34 µm per pixels, while predictions of U-Net$_{synth}$ on the H&E Cryo dataset were obtained using a pixel resolution of 0.6615 µm per pixels. Additionally, for the U-Net$_{real}$ and U-Net$_{synth}$ predictions the same post processing was used. The expert survey as well as all manual annotation on the H&E Cryo dataset including the ROI definitions was carried out using Cytomine [58], an open-source collaborative web environment.

**Statistical analyses**

To assess potential significant differences between the ranked U-$Net_{real}$ and U-$Net_{synth}$ predictions within the expert survey, we performed a paired t-tests with a significance level of $\alpha = 0.05$ for each category "Segmentation", "Shape Filter" and "Connective Tissue" ($n=156$ ROIs per category) and for each expert (P1 and P2), respectively. The t-tests were computed using SciPy [59] (version 1.4.1), a Python scientific computing library.

**Post Processing**

A custom Fiji (version v1.53f51) toolset was used to post process the model predictions in a fully automated manner. It is publicly available under https://github.com/OliverAust/HE_Muscle_Seg. First, the predicted probability maps were scaled to a range of [0, 255] and an intensity threshold of 210 was set to obtain the raw binary segmentation masks. Next, the objects were further filtered using a minimum object size of 150 µm. Additionally, to exclude apparent non-organic shaped fibers from the analysis, objects below a circularity threshold of 0.1 were removed. The resulting binary mask is referred to as "post processed" in the following. For H&E FFPE image evaluation only these post processed images were considered, while for H&E Cryo images the following 3 post processing steps were additionally performed.

*Segmentation Mask*

Masks of the whole muscle slice were generated. This was realized by dilating the post processed images ten times, followed by ten erosion operations. To remove muscle mask errors, the Fiji [42] function "Area Opening" from the plugin MorphoLibJ [60] was applied with a maximum size of 3,500 pixels². To finalize muscle mask generation, a size filter of 1,500,000 µm² was applied. The muscle masks were then applied on the post processed images to remove any muscle fragments outside the muscle slice. The resulting images were used in the evaluation of the segmentation.

*Shape Filtering*

A shape filter was used on the post processed image to generate additional images with pathologically relevant fibers only. A minimum circularity of 0.35 and maximum circularity of 0.95 were used for shape filtering. Muscle masks were also generated from shape filtered image as they are required for the connective tissue generation.

*Connective Tissue*

Creating the difference between the muscle mask and the minimal threshold, or the shape filtered muscle mask and shape filtered images yields the connective tissue of the respective image. To evaluate the connective tissue, a selection was created from the respective muscle mask and then applied to the connective tissue. The connective tissue fraction and total area was measured inside the muscle mask area. The post processing, shape filtering and connective tissue workflow is visualized and discussed in detail in Supp. Fig. 2-4.

### Data availability

The data that support the plots within this manuscript and other findings of this study are available from the corresponding author upon request.

### Code availability

Upon publication the Fiji toolset implementing the post processing procedure will be freely available under https://github.com/OliverAust/HE_Muscle_Seg.

**Acknowledgements**

The research leading to these results has received funding from the European Research Council (ERC) (Grant agreement No. 810316). The authors thank Carolin Berwanger, Institute of Aerospace Medicine, German Aerospace Center (DLR), Cologne, Germany, for essential support in maintenance and dissection of desmin knock-out mice (DKO Hom). We thank M. Gunzer for critical reading of the manuscript.

**Disclosures**

A patent application has been filed by L.M. for the parametric generation of synthetic biomedical image data as training data in the context of AI-based biomedical image analysis.

**Author contributions**

L.M.: Conceived the idea for the simulation; designed and implemented the simulation; generated the synthetic data; trained deep learning models; helped with the implementation of the Fiji toolset; contributed to study design; generated and evaluated results; wrote manuscript.

O.A.: Conceived the idea for the muscle fiber segmentation task; implemented the Fiji toolset; contributed to study design; generated and evaluated results; wrote manuscript.

J.A.A., P.B., M.P., K.P.Z.: Generation of dataset A (H&E FFPE), including sample preparation and imaging; labeling of images for U-$\text{Net}_{\text{real}}$ training; helped improving the manuscript and Fiji toolset.

G.K: Organization between departments; planning of dataset A (H&E FFPE) experiments; commented on manuscript.

S. U., G. S.: Organization between departments; feedback on Fiji toolset; commented on manuscript.

C.S.C.: Breeding, maintenance, genotyping, and dissection of desmin knock-out mice; preparation of cryo-preserved skeletal muscle samples; participation in manuscript writing, discussed the results and suggested improvements.

R.S.: Histopathological analysis of murine desmin knock-out skeletal muscle sections; participation in manuscript writing; participation in data analysis; discussed the results and suggested improvements.

C.H.: Annotation of regions of interest in H&E-stained sections from murine desmin knock-out skeletal muscle; digitalization of histological slides.

S.J.: Annotations of regions of interest in H&E-stained sections from murine desmin knock-out skeletal muscle; contributed to the study design; participation in data analysis; discussed the results and suggested improvements.

A.M.: Helped with the study design and the data analysis; commented on the manuscript.

A.G.: Conceived of and supervised the study; wrote the manuscript together with L.M., O.A. and with the help of J.A.A., P.B., M.P., K.P.Z., G.K., S.U., G.S. C.S.C., R.S., C.H., S.J., and A.M..

All authors contributed to discussions and writing of the manuscript.

## Competing interests

RS serves as a consultant for MIRA Vision Microscopy GmbH. The other authors declare no competing interests.

## Supplementary Figures

**a** Parametric synthetic image variations of H&E-stained skeletal muscle fibers

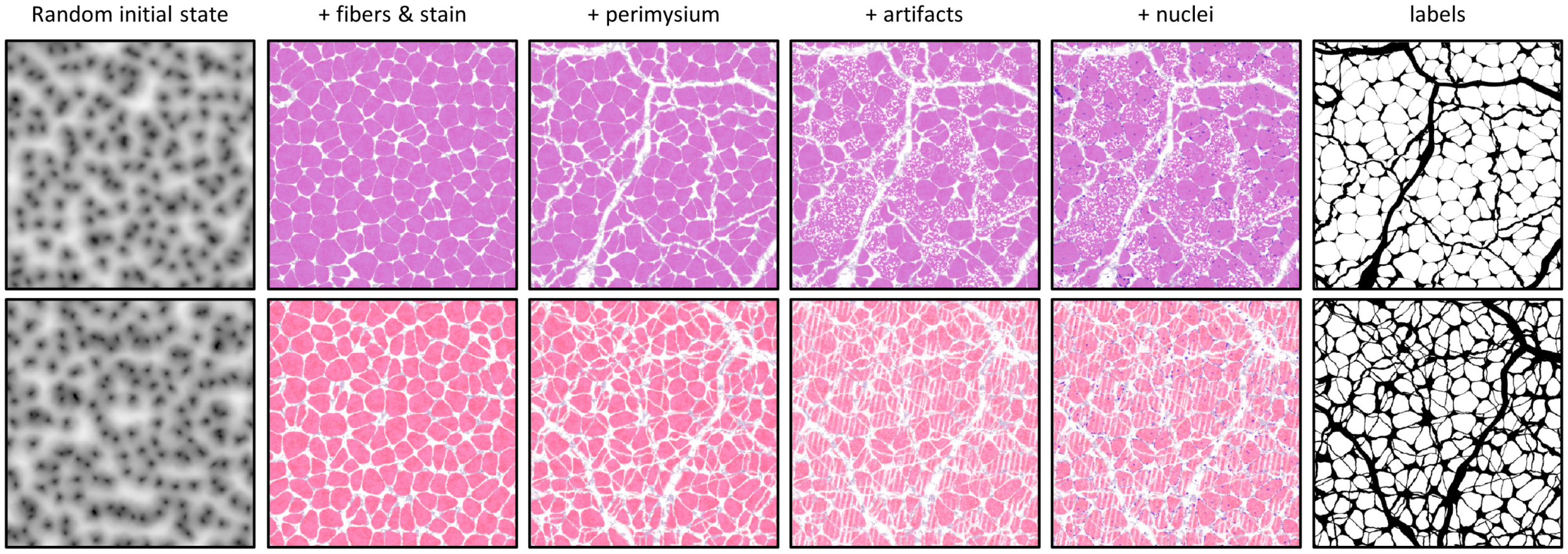

**b** Example synthetic training images

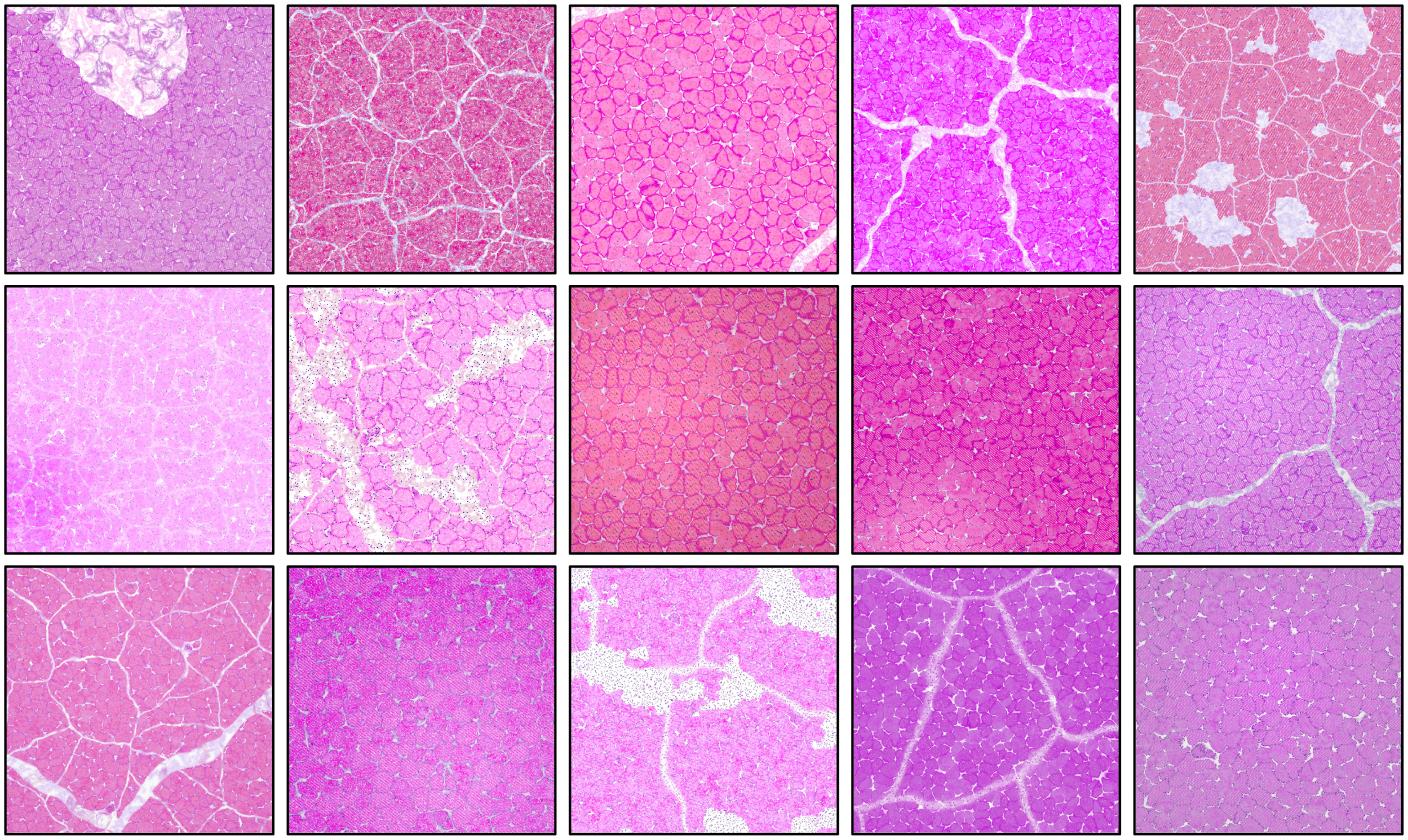

**Supp. Fig. 1 | Parametric synthetic image variations of H&E-stained skeletal muscle fibers. a,** Generation of realistic muscle fibers images. Based on a random initial state, randomized fibers, staining, perimysium, artifacts, and nuclei are added to obtain realistic renders while guaranteeing perfect segmentation labels with no annotation noise. **b,** Example realistic and non-realistic synthetic training images that were randomly sampled from the parametric image generation pipeline.

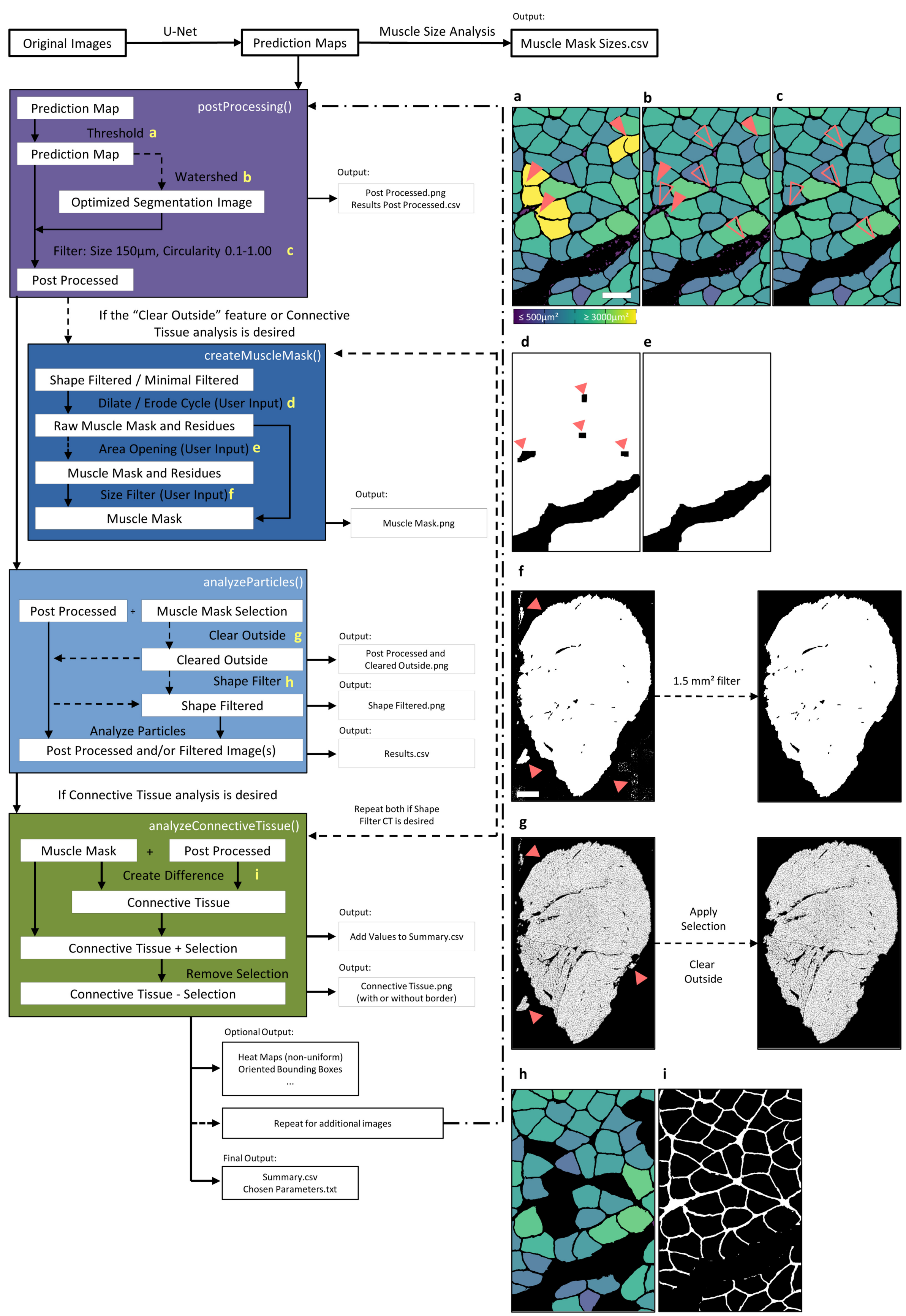

Original Images
U-Net
Prediction Maps
Muscle Size Analysis
Output:
Muscle Mask Sizes.csv
postProcessing()
Prediction Map
Threshold a
Prediction Map
Watershed b
Optimized Segmentation Image
Filter: Size 150µm, Circularity 0.1-1.00 c
Post Processed
Output:
Post Processed.png
Results Post Processed.csv
If the "Clear Outside" feature or Connective Tissue analysis is desired
createMuscleMask()
Shape Filtered / Minimal Filtered
Dilate / Erode Cycle (User Input) d
Raw Muscle Mask and Residues
Area Opening (User Input) e
Muscle Mask and Residues
Size Filter (User Input) f
Muscle Mask
Output:
Muscle Mask.png
analyzeParticles()
Post Processed
+
Muscle Mask Selection
Clear Outside g
Cleared Outside
Shape Filter h
Shape Filtered
Analyze Particles
Post Processed and/or Filtered Image(s)
Output:
Post Processed and Cleared Outside.png
Output:
Shape Filtered.png
Output:
Results.csv
If Connective Tissue analysis is desired
Repeat both if Shape Filter CT is desired
analyzeConnectiveTissue()
Muscle Mask
+
Post Processed
Create Difference i
Connective Tissue
Connective Tissue + Selection
Remove Selection
Connective Tissue - Selection
Output:
Add Values to Summary.csv
Output:
Connective Tissue.png (with or without border)
Optional Output:
Heat Maps (non-uniform)
Oriented Bounding Boxes
...
Repeat for additional images
Final Output:
Summary.csv
Chosen Parameters.txt
a
b
c
≤ 500µm²
≥ 3000µm²
d
e
f
1.5 mm² filter
g
Apply Selection
Clear Outside
h
i

**Supp. Fig. 2 | Visualization of the Post Processing Workflow Diagram:** The bold letters next to certain post processing steps in the diagram are represented by the image on the right side with associated bold letter. Functions of the Fiji Macro Code are indicated by a box with the function name on the top right to facilitate changes to the code. **a,** The original images are converted to prediction maps with the U-Net and subsequently a threshold is applied. Note that since fibers are merged the resulting area of that object is large and thus the fibers appear yellow. Higher thresholds can cause a higher area error but can also cause a smaller segmentation error. Three examples for segmentation errors are indicated by pink filled arrowheads. **b,** An optional watershed can segment such fibers efficiently as again indicated by the filled pink arrowheads. Additionally, the $\text{U-Net}_{synth}$ misinterprets some small tissue regions as muscle fibers (pink open arrowheads). **c,** These can be removed by a small size and circularity filter. The resulting images are referred to as "post processed" in this publication. To further improve analysis results and as a basis for connective tissue analysis a mask of the whole muscle section will be created. This is realized by several (e.g., 10) dilate operations that cause fiber-fiber gaps to be closed but also cause an enlargement of every structure. **d,** To keep the original outline of the whole muscle section, the same amount of erode operations are applied to the image. Completely filled holes and spaces are unaffected by the erode operations. Depending on the amount of dilate and erode operations applied, small holes will remain (pink filled arrowheads). **e,** These can be removed by inverting the image, using the area opening operation by MorphoLibJ [60] and inverting the image again. **f,** Finally, for whole slide images, a size filter must be applied to remove any unwanted muscle tissue fragments around the main muscle sample. **g,** The resulting mask of the whole muscle section is referred to as 'muscle mask'. This mask can be applied to the post processed image to remove any muscle tissue fragments outside the main sample that were left after the shape filter. **h,** Before the remaining objects in the image are analyzed and results are saved it is possible to apply a shape filter. The last step carried out is the analysis of the connective tissue. **i,** First, to create connective tissue images the difference between the muscle mask and the post processed image is generated. Since the gaps between the fibers are closed in the muscle mask image while they are still open in the post processed image, the difference of both images will be equal to fiber-fiber gaps. Local Mean thickness and the muscle mask selection is then used as the base to calculate connective tissue to fiber ratio. Shown images are for illustration purpose only. Scale bar **a**, 50 µm; Scale bar **f**, 500 µm.

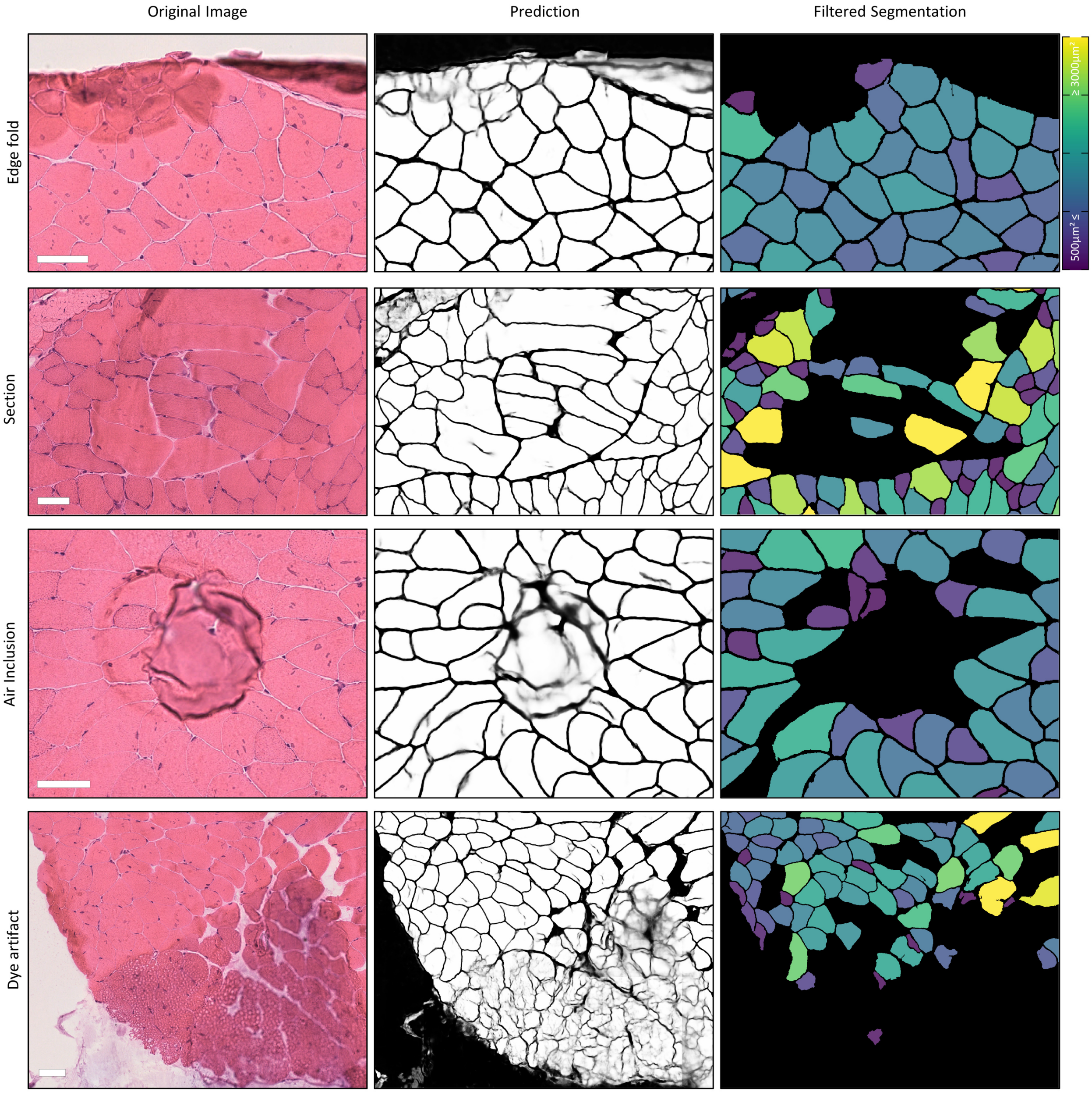

**Supp. Fig. 3 | Artifact Handling during Post Processing.** Sample preparation artifacts such as air inclusion and folded tissue (original unaltered images left column) can be misinterpreted by the U-Net as muscle fibers (predictions middle) since the fibers in artifacts still share many properties of normal muscle fibers such as form and coloring. As it is most likely that an inclusion of these fibers in the analysis is not desired, it is crucial to be able to remove them. With the use of shape filters such as circularity it is possible to remove most affected fibers (filtered segmentation right). Not all fibers can be removed as some artifacts have very fiber similar shapes. Since the type, occurrence rate and similarity to desired disease model observations will heavily depend on each individual project, these parameters can be adjusted accordingly in the available macro. The filter parameters used: Folded tissue: threshold set at 120, 15 maximum watershed, circularity filter of 0,45-1,00. Oblique and longitudinally oriented fibers within a cross section: threshold set at 140, 45µm maximum feret diameter filter, 0,4-1 circularity filter. Air inclusion: threshold set at 140, 0,45 – 1 circularity. Stain spilling: threshold set at 140, 0,4 – 1 circularity filter. Shown images are for illustration purpose only. Scale bars, 50µm.

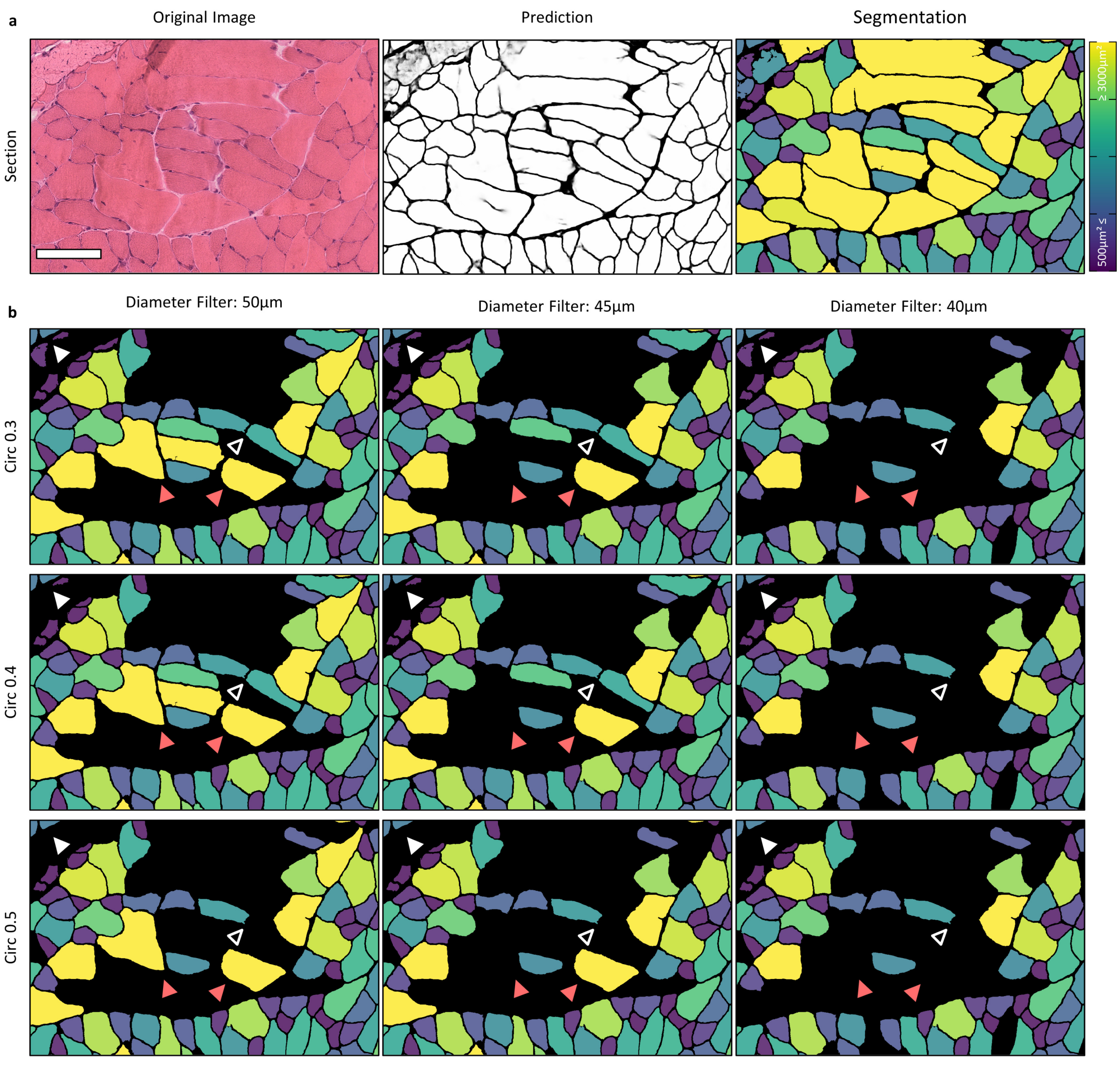

**Supp. Fig. 4 | Visualization of the different shape filter parameters. a,** shows the original image, the unprocessed prediction and the final segmentation. The depicted area of the cross section includes many oblique and longitudinally oriented fibers that users might want to exclude from their statistical analysis. **b,** shows the effect of the indicated shape filter parameter combinations in regard to the excluded fibers. It can be seen that some fibers can be excluded by both parameters (indicated with the white open arrowheads). Both a maximum Feret diameter and a minimal circularity filter can exclude these fibers. However, it can also be seen that some fibers can only be filtered by one of the parameters. For example, some fibers are only removed by setting a maximum Feret diameter while a circularity filter does not have any effect (magenta filled arrowheads). In contrary, smaller fibers originating from neural network misinterpretation (white filled arrowheads) are only removed by setting a circularity filter. Shown images are for illustration purpose only. Scale bar, 100µm.

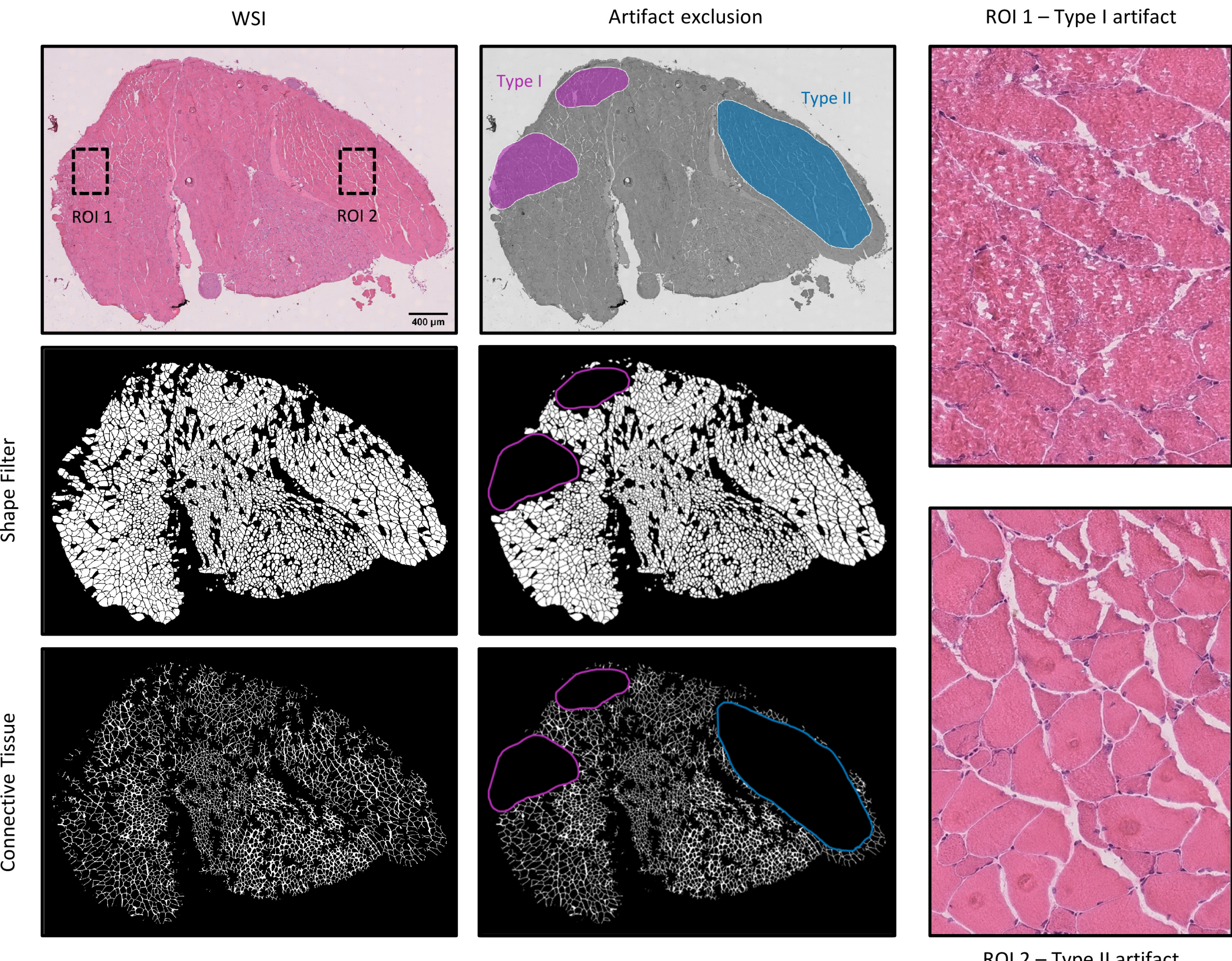

**Supp. Fig. 5 | Exlusion of regions with severe artifacts from the analysis.** For an accurate analysis of the H&E Cryo dataset, regions with marked freezing artifacts (Type I) or artificial loosening (Type II) were manually annotated. Fibers and connective tissue of the “Shape filter” and “Connective Tissue” that lied within such regions were automatically removed from the masks and excluded from the dataset analysis.

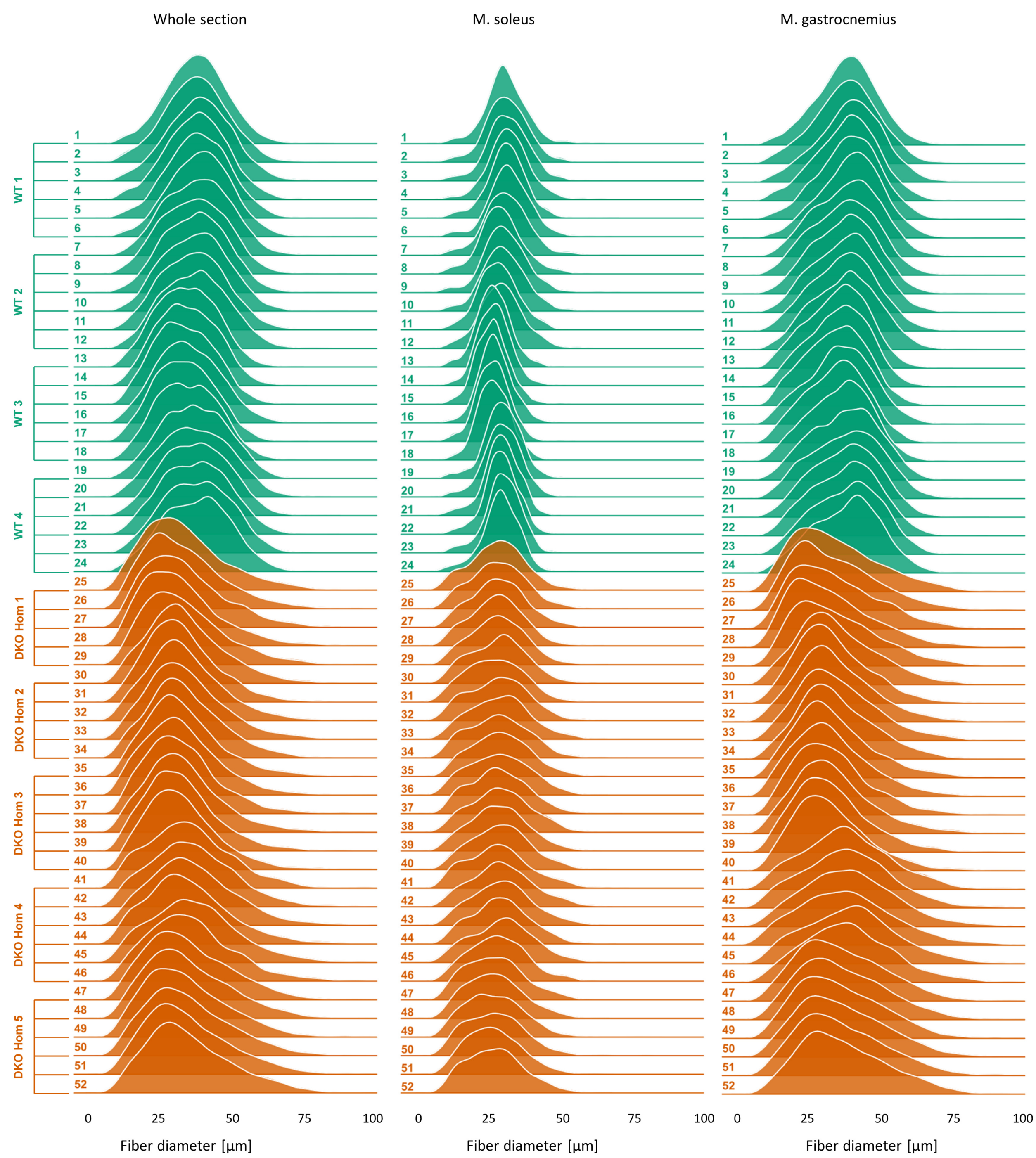

**Supp. Fig. 6 | Kernel density estimation (KDE) plots of the fiber diameter distributions of each WT (green) and DKO Hom (orange) WSI section.** The KDEs are visualized separately either for the whole section, the M. soleus only or the M. gastrocnemius only. Additionally, the sections are grouped by the tissue samples, wild-type siblings WT 1 – WT 4 and desmin knock-out animals DKO Hom 1 – DKO Hom 5. The plots visualize an increased variance of fiber diameter distribution and a decrease of mean fiber diameter in desmin knock-out skeletal muscle.